\documentclass[11pt,sort&compress]{elsarticle}

\usepackage{amssymb}
\usepackage{amsmath}
\usepackage{graphicx}
\usepackage{tabularx}
\usepackage{cite}
\usepackage{color}
\usepackage{fullpage}
\usepackage{subfigure}
\usepackage{subfigmat}
\usepackage{etoolbox}
\usepackage{url}
\usepackage{float}
\usepackage{bm}
\usepackage[dvipsnames]{xcolor}
\usepackage{placeins}
\usepackage{siunitx} 
\usepackage[nameinlink,capitalize]{cleveref}
\journal{International Journal of Multiphase Flow}

\begin{document}

\begin{frontmatter}

\title{
Effect of stand-off distance and spatial resolution on the pressure impact of near-wall vapor bubble collapses
}

\author{Theresa Trummler}
\ead{theresa.trummler@tum.de}
\author{Steffen J. Schmidt}
\author{Nikolaus A. Adams}

\address{
  Chair of Aerodynamics and Fluid Mechanics, Technical University of Munich \\ 
  Boltzmannstr.\ 15, 85748 Garching bei M\"unchen, Germany}

\begin{abstract}
    We consider the collapse behavior of cavitation bubbles near walls under high ambient pressure conditions. Generic configurations with different stand-off distances are investigated by numerical simulation using a fully compressible two-phase flow solver including phase change. The results show that the stand-off distance has significant effects on collapse dynamics, micro-jet formation, rebound, and maximum wall pressure. A relation between cavitation induced material damage and corresponding collapse mechanisms is obtained from pressure-impact data at the wall. 
    
    We analyze the resolution dependence of collapse and rebound and the observed maximum pressure distributions. The comparison of the results on six different grid resolutions shows that main collapse features are already captured on the coarsest resolution, while the peak pressures are strongly resolution dependent.

\end{abstract}


\end{frontmatter}
\section{Introduction}
\label{sec:Intro}

When a cavitation bubble collapses in the vicinity of a wall, a wall-directed jet is formed~\citep{naude1961mechanism,Plesset:1971hu}, and intense pressure waves are emitted from the collapse site~\citep{hickling64, Fujikawa:1980jj}. Jet and pressure waves cause considerable material damage, representing a serious problem in marine applications, hydraulic power generation and injector components~\citep{arndt1981cavitation,asi2006failure,Franc:2004fu}. On the other hand, cavitation bubbles are also exploited to benefit from the intense localized pressure waves and energy focusing, as e.g. in biomedical applications, such as urinary stone ablation~\citep{Pishchalnikov:2003xv} and drug delivery~\citep{coussios2008applications}, or new technologies, such as surface-cleaning~\citep{Ohl:2006fs,Reuter:2017bu}, micro-pumps in microfluidics~\citep{dijkink2008laser} and water treatment~\citep{Zupanc:2012ex}. In this paper, the cavitation dynamics and damage potential of near-wall bubble collapses are numerically studied in detail. Furthermore, we quantify the effect of the grid resolution on the numerical results, which is relevant for engineering predictions. 

Early studies by \citet{benjamin66,Plesset:1971hu, hickling64, naude1961mechanism,Shima:1977df} already reported on the damage potential of collapses in the vicinity of walls and found an aspherical behavior leading to the formation of a jet. Later, \citet{Shima:1983we, Shima:1984vw,Tomita:1986gy,Philipp:1998eg} experimentally examined the erosion potential of single bubble collapses and showed that the distance to the wall is decisive for the damage potential. However, despite numerous studies, the role of the jet and the shock wave for the damage is still not fully understood~\citep{dular2019high}. The dynamics of aspherical collapses have been studied by e.g. \citet{Lindau:2003oct, brujan2002final, brujan2005jet, Kroninger:2009dz} and more recently by \citet{Supponen:2017sw, Supponen:2016jna,Supponen:2018tv}. Over the last decade, compressible numerical simulations have started to complement experimental studies allowing for new insights with their time-resolved flow field data. For example, \citet{Johnsen:2009cua, Lauer:2012jh} numerically studied the collapse behavior and erosion potential of bubbles in the vicinity of walls. While most numerical studies focus on the first collapse \citep[e.g.][]{Johnsen:2009cua, Beig:2018ga,Pishchalnikov:2019,Trummler:2019ar}, \citet{ochiai2011numerical}, and more recently \citet{sagar2020dynamics}, also evaluated the erosion potential of the second collapse by considering phase change.

We revisit the dynamics and damage potential of aspherical bubble collapses and investigate the underlying mechanisms. For this purpose, we perform compressible, high-resolution, 3-D simulations of collapsing vapor bubbles with varying stand-off distances to a wall. Following previous studies \citep[e.g.][]{Beig:2018ga,Lauer:2012jh, Kyriazis:2018dz,Trummler:2019ar}, we consider a driving pressure of 100 bar, which is representative for high-pressure applications such as pumps~\citep{bohner01} and injector components~\citep{Payri:2009bl}. 
Our goal is to determine where the highest pressures occur and to investigate the relevant processes jet impact, collapse and rebound and their effects in detail. For the modeling of the two-phase flow, we employ a homogeneous mixture model with a thermodynamic equilibrium cavitation model \citep{schmidt2006compressible,Schnerr:2008jja}. By considering phase change we avoid possible effects of non-condensable bubble content on jet velocity, maximum pressure and rebound~\citep{Pishchalnikov:2019,Tinguely:2012wo,Trummler:2018ww}. Furthermore, the modeling of evaporation allows us to capture the rebound in post-collapse low-pressure regions. 

In numerical simulations, the result can be affected by a variety of modeling parameters, such as the grid resolution~\citep{Mihatsch:2015db}, the numerical scheme~\citep{Egerer:2016it,schmidmayer19}, the thermodynamic modeling of the liquid~\citep{Kyriazis:2018dz}, and the modeling of the bubble content (see above). The grid resolution is one of the most important aspects and determines the minimum resolvable bubble radius (if non-condensable bubble content is present)~\citep[e.g.][]{Beig:2018ga}, the form and intensity of the rebound~\citep[e.g.][]{schmidmayer19}, and especially the intensity of the pressure peaks~\citep[e.g.][]{schmidt2014assessment,Mihatsch:2015db,Mihatsch:2017diss}. Since there are no detailed previous studies on grid dependence of aspherical bubble collapses, we consider six different grid resolutions and determine the grid dependence of the rebound and the pressure peaks. Based on the literature, a grid resolution of about 100 cells $c$ over the maximum radius $R_0$ can be considered as state of the art \citep{Lauer:2012jh}. The highest reported resolutions are $200\,c/R_0$ for 3-D \citep{Beig:2018ga} and $1600\,c/R_0$ for axisymmetric 2-D simulations \citep{Koch:2017bd}. The results presented here of resolutions of $240\,c/R_0$ and $320\,c/R_0$ represent the highest grid resolution used so far for 3-D simulations. 

The paper is structured as follows. In \cref{sec:Methods}, we describe the physical model and numerical method. \Cref{sec:NumSetup} presents the numerical set-up and the considered configurations. Then, the collapse behavior of the bubble, its pressure impact on the wall and the maximum pressure distribution are analyzed for different stand-off distances in \cref{sec:stand_off}. In \cref{sec:grid}, the effect of the grid resolution on the collapse dynamics and the pressure impact is evaluated by comparing results for six different grid resolutions. \Cref{sec:Conclusion} summarizes the paper. 

\section{Physical model and numerical method}
\label{sec:Methods}

\subsection{Governing Equations}
\label{subsec:GovEq}
We solve the fully compressible Navier-Stokes equations in conservative form 
  \begin{equation}
    \partial_{t}\boldsymbol{U}+\nabla \cdot [ \boldsymbol{C}(\boldsymbol{U})+\boldsymbol{S}(\boldsymbol{U})]=0\,. 
    \label{eq:NS}
  \end{equation}
The state vector $\boldsymbol{U}=[\rho , \, \rho \boldsymbol{u} ]^T$ is composed of the conserved variables density $\rho$ and momentum $\rho \boldsymbol{u}$. Due to the assumed barotropic modeling ($p=p(\rho)$), the energy equation can be omitted. The convective fluxes $\boldsymbol{C}(\boldsymbol{U})$ and the flux contributions due to pressure and shear $\boldsymbol{S}(\boldsymbol{U})$ read
\begin{equation}
  \boldsymbol{C}(\boldsymbol{U})=
      \boldsymbol{u}
        \begin{bmatrix} 
         \rho\\ 
         \rho\boldsymbol{u}\\ 
      \end{bmatrix}
      \quad
      \mathrm{and}
      \quad
      \boldsymbol{S}(\boldsymbol{U})=
      \begin{bmatrix} 
      0\\ 
       p \boldsymbol{I}-\boldsymbol{\tau}\\ 
      \end{bmatrix},
         \label{eq:NS_Basis}
  \end{equation}  
with the velocity $\boldsymbol{u}$, the static pressure $p$, the unit tensor $\boldsymbol{I}$, and the viscous stress tensor $\boldsymbol{\tau}$
\begin{equation}
 \boldsymbol{\tau}=\mu(\nabla \boldsymbol{u}+(\nabla \boldsymbol{u})^{T}-\frac{2}{3}(\nabla \cdot\boldsymbol{u})\boldsymbol{I}),
 \label{eq:tau}
\end{equation}
where $\mu$ is the dynamic viscosity.

\subsection{Thermodynamic Model}
\label{subsec:ThermoModel}

\citet{schmidt2006compressible} and \citet{Schnerr:2008jja} formulated a cavitation model based on the assumption that the liquid and gas phase of a cavitating liquid are in thermal and mechanical equilibrium. This model has been successfully validated for spherical collapse with analytical solutions \citep[e.g.][]{Egerer:2016it,Trummler:2018ww} and aspherical collapse with the collapse behavior and pressure data of \citet{Lauer:2012jh} by e.g. \citet{Pohl:2015keb, Oerley:2016diss,Koukouvinis:2016ir}. Apart from that, it has been extensively validated for cavitating nozzle flows~\citep{Egerer:2014wu, Orley:2015kt, Trummler:2018AAS, Trummler:2020if, budich2018jfm} and numerical cavitation erosion prediction~\citep{Mihatsch:2015db,schmidt2014assessment} with experimental data. 

Due to the thermodynamic equilibrium assumption for the cavitation model, finite-rate mass transfer terms are avoided. The liquid starts to cavitate if the pressure drops beneath saturation pressure 
   \begin{equation}
    p<p_\mathrm{sat}
    \label{eq:p}
  \end{equation}
and then a liquid-vapor mixture is present. For instantaneous phase change in local thermodynamic equilibrium, the densities of liquid and vapor are their saturation densities $\rho_{\mathrm{sat,}l}$, $\rho_{\mathrm{sat,}v}$. Thus, the vapor volume fraction $\alpha$ is given by the density of the liquid-vapor mixture $\rho$ as 
  \begin{equation}
    \alpha=\frac{\rho_{\mathrm{sat,}l}-\rho}{\rho_{\mathrm{sat,}l}-\rho_{\mathrm{sat},v}}. 
  \end{equation}
For water at reference temperature $\mathrm{T}=\SI{293.15}{K}$, the corresponding values are $p_\mathrm{sat}=\SI{2340}{Pa}$, $\rho_{\mathrm{sat,}l}=\SI{998.1618}{kg/m^3}$ and $\rho_{\mathrm{sat,}v}= \SI{17.2e-3}{kg/m^3}$. 

Different equations of state for the pure liquid and the liquid-vapor mixture are employed. The pure liquid is modeled using a modified Tait equation
  \begin{equation}
    p=B\left(\left(\frac{\rho}{\rho_\mathrm{sat,l}}\right)^{N}-1\right)+p_\mathrm{sat} \quad \mathrm{for} \quad \rho\geq\rho_{\mathrm{sat,}l},
    \label{eq:Tait}
  \end{equation}
with the fluid-specific parameters $N= 7.15$ and $B= \SI{3.3e8}{\pascal}$ \citep{Schnerr:2008jja, Mihatsch:2015db,Lauer:2012jh}. 

The equation of state for the two-phase region ($p<p_\mathrm{sat}$) is derived from the definition of the isentropic speed of sound $c$
  \begin{equation}
    c=\sqrt{\frac{\partial p}{\partial \rho}\bigg|_{s=\mathrm{const.}}}\;\;.
    \label{eq:c0}
  \end{equation}
Assuming that phase change takes place in equilibrium, we use the equilibrium speed of sound $c_\mathrm{eq}$ \citep{Franc:2004fu} 
  \begin{equation}
    \frac{1}{\rho\,c_\mathrm{eq}^2}\approx \frac{\alpha}{\rho_v\,c_v^2}+\frac{1-\alpha}{\rho_l\,c_l^2}+\frac{(1-\alpha)\,\rho_l \,c_{\mathrm{p,}l}\,T}{(\rho_v\,L)^2}, 
    \label{eq:c_eq}
  \end{equation}
where $c_{\mathrm{p},\varphi}$ is the specific heat capacity, $c_{\varphi}$ the speed of sound of the component $\varphi=\{v,l\}$, and $L$ the latent heat. 
Integration from saturation conditions yields into 

  \begin{equation}
    p=p_\mathrm{sat}+\frac{1}{C_1}\,log\left(\frac{\rho}{C_1+C_2\,\rho}\right) \quad \mathrm{for}\quad \rho<\rho_{\mathrm{sat,}l}, 
    \label{eq:p_lv_1}
  \end{equation}
with the fluid-specific constants $C_1$ and $C_2$
  \begin{equation}
    C_1=C_3-\frac{C_4 \,\rho_l}{\rho_v-\rho_l}, 
    \quad 
    C_2=\frac{C_4}{\rho_v-\rho_l}, 
    \label{eq:c1_c2}
  \end{equation}
where $C_3$ and $C_4$ are defined by 
 \begin{equation}
    C_3=\frac{1}{\rho_l\, c_l^2}+\frac{\rho_l \,c_{\mathrm{p,}l} \,T}{(\rho_v L)^2}, 
    \quad
    C_4=\frac{1}{\rho_v \,c_v^2}-\frac{1}{\rho_l\, c_l^2}+\frac{T(\rho_v \,c_{\mathrm{p,}v}-\rho_l \,c_{\mathrm{p,}l})}{(\rho_v \,L)^2}. 
    \label{eq:c3_c4}
  \end{equation}
We have used the following values: $c_l=\SI{1482.2}{m/s}$, $c_v=\SI{423.18}{m/s}$, $\rho_l=\rho_{\mathrm{sat,}l}$, $\rho_v=\rho_{\mathrm{sat,}v}$, $c_{\mathrm{p,}l} =\SI{4.18}{kJ/kg K}$, $c_{\mathrm{p,}v} =\SI{1.91}{kJ/kg K}$, and $L= \SI{2.45}{MJ/kg}$. 

Viscous effects are considered in our simulations. The dynamic viscosity of the mixture is determined by 
  \begin{equation}
    \mu=\alpha\cdot\mu_v+(1-\alpha)\cdot\mu_l, 
    \label{eq:muLiqMix}
  \end{equation}
with $\mu_l = \SI{1.002e-3}{\pascal \second}$ and $\mu_v = \SI{9.272e-6}{\pascal \second}$. Surface tension is neglected in our model as it is only significant at a bubble radius of $R<3\,S/\Delta p$~\citep{Franc:2004fu}, which is at the considered conditions (driving pressure $\Delta p=\SI{e7}{\pascal}$, surface tension $S = \SI{0.072}{N/m}$) on the negligible order of $O(\SI{e-8}{m})$. 

\subsection{Numerical Method}
The thermodynamic model is embedded into a density-based fully compressible flow solver with a low-Mach-number-consistent flux function, see \citet{Schmidt:2015wa}. For the reconstruction at the cell faces an upwind biased scheme is used, where the velocity components are reconstructed with the up to third-order-accurate limiter of \citet{Koren:1993} and the thermodynamic quantities $\rho$, $p$ with the second-order minmod slope limiter of~\citet{Roe:1986}.

Time integration is performed with an explicit second-order, 4-step low-storage Runge-Kutta method \citep{Schmidt:2015wa}. 


\begin{figure}
  \centering
  \subfigure{\includegraphics[width=0.8\linewidth]{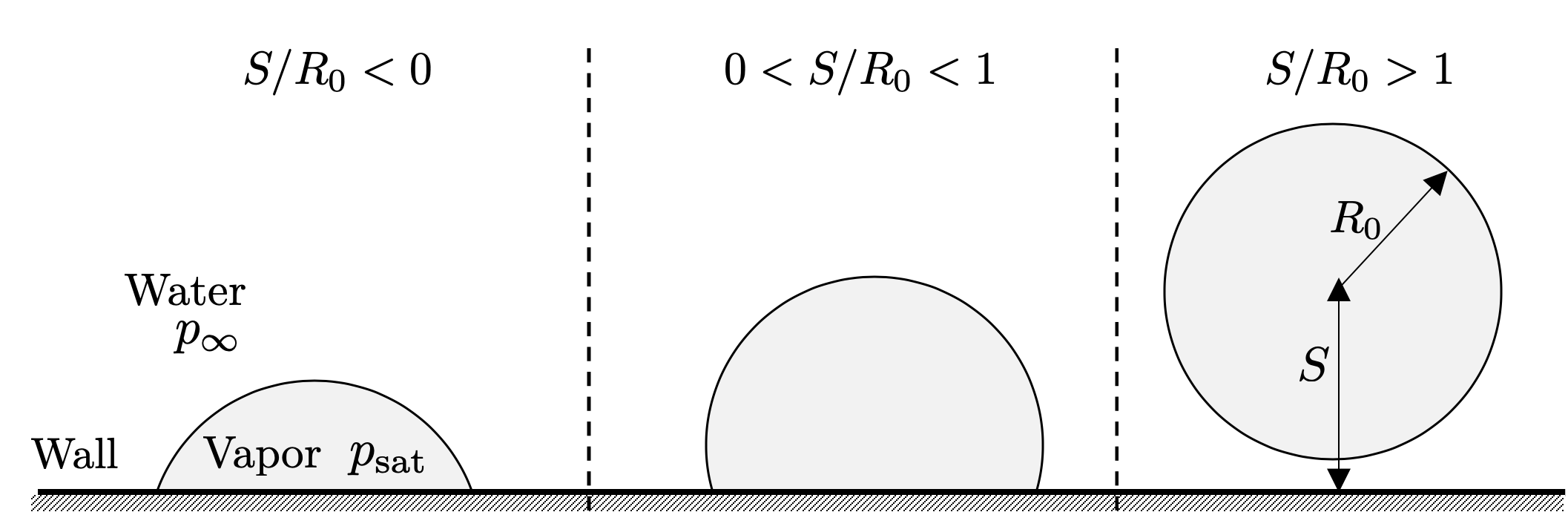}}
  \caption{Considered configurations.}
  \label{fig:setup}
\end{figure}

\section{Set-up}
\label{sec:NumSetup}

We consider a vapor bubble with initial radius $R_0=\SI{4e-4}{\metre}$ and varying stand-off distances $S$ from the wall as shown in \cref{fig:setup}. The bubble is surrounded by water with a driving pressure difference of $\Delta p=\SI{e7}{\pascal}$, and the pressure field is initialized as 
\begin{equation}
  p(\hat{r},t=0) = p_\mathrm{sat}+\Delta p \, \frac{\hat{r}-R_0}{\hat{r}} 
  \;\; \text{for} \;\; \hat{r} > R_{0},
  \label{eq:p_liq}
\end{equation}
where $\hat{r}$ is the radial coordinate with its origin at the bubble center. This initialization matches the pressure distribution predicted by the Rayleigh equation for the Besant problem~\citep{brennen95,besant59}. For the configurations considered, it provides a valid approximation of the realistically evolving pressure field and suppresses the formation of spurious pressure waves due to pressure jumps at the bubble interface. \citet{Rasthofer:2019co} have demonstrated that such an approximation evolves towards an exact solution of the Besant problem within a very short time. 

Our initial conditions are chosen to facilitate comparison with other studies, including \citet{Beig:2018ga,Lauer:2012jh}, and, most importantly, to provide well-defined and easily reproducible conditions. To assess the effect of the grid resolution on the collapse behavior and the wall pressure impact, a resolution-independent initialization procedure is essential. Initial conditions similar to experimental ones can be obtained by including the bubble growth phase in the simulation, see e.g. \citet{Lauterborn:2018cq, Lechner:2019fp, Lechner:2020jet}.

\begin{figure}
  \subfigure[]{\includegraphics[trim={250 0 0 0},clip,height=6cm]{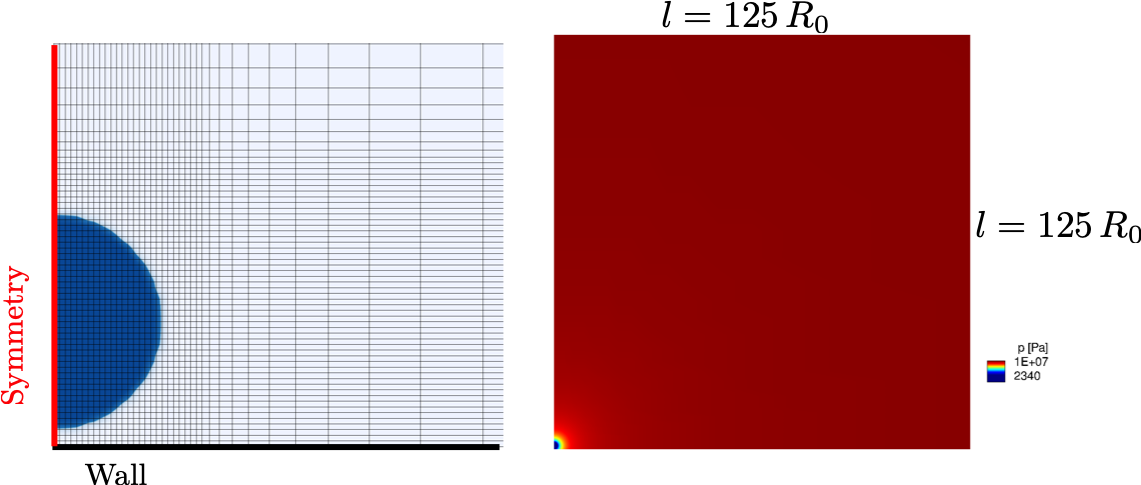}}
  \hspace{1cm}
  \subfigure[]{\includegraphics[trim={0 0 300 0},clip,height=6cm]{figure02.png}}
  \caption{Numerical set-up and grid. (a) Domain and initial pressure distribution on the midplane. (b) Grid in the near bubble region for $S/R_0=1.1$ for the coarsest resolution of $20\,c/R_0$.}
  \label{fig:grid}
\end{figure}

\Cref{fig:grid} shows the numerical set-up and the computational grid. Taking advantage of symmetry, only a quarter of a bubble is simulated. The bubble is put in the center of a 3-D domain with a length of $250\,R_0$ in wall-parallel direction and of $125\,R_0$ in wall-normal direction. The grid is equally spaced with a defined number of cells per initial radius ($N_c/R_0$) in the near bubble region, which is until $\hat{r} = 1.25\,R_{0}$ in wall-parallel direction and $\hat{r} = 2.5\,R_{0}$ in wall-normal direction, and is progressively stretched farther from the bubble. The highest grid resolution is $320\,c/R_0$ with a total of 670 million cells. A constant CFL number of $1.4$ is used, which corresponds to a time step of $\Delta t \approx \SI{e-10}{\second}$ at the highest grid resolution. The total simulation time is adapted to capture the second collapse and is about $\SI{8e-6}{\second}$. 

Results are presented in a non-dimensional form. Therefore the time scale is normalized with 
\begin{gather}
  t^{\ast} = R_0 \sqrt{\frac{\rho_l }{ \Delta p}},  
  \label{e:collapse}
\end{gather}
which is an estimate of the collapse time of a near-wall bubble collapse \citep{Plesset:1971hu}. The wall has a retarding effect on the collapse and thus $t^{\ast}$ is longer than the Rayleigh collapse time for spherical collapses ($t_\mathrm{Rayleigh}=0.915\,t^{\ast}$)~\citep{Rayleigh:1917}. Velocity and pressure are normalized with
\begin{gather} 
  u^{\ast} = \sqrt{\frac{\Delta p}{\rho_l}}, 
  \quad \text{and} \quad
  p^{\ast} = c_l \sqrt{\rho_l \Delta p}.
\end{gather}
Note that $p^{\ast}$ corresponds to a water hammer pressure $p_{\mathrm{WH}}$ induced by the velocity $u^{\ast}$ as
\begin{equation}
  p_{\mathrm{WH,}u^{\ast}}= \rho_l c_l u^{\ast}=\rho_l c_l \sqrt{\Delta p/\rho_l}=c_l \sqrt{\rho_l \Delta p} \equiv p^{\ast}. 
  \label{eq:pwh}
\end{equation}
Additionally, the employed expression for $p^{\ast}$ can be related to the scaling found by \citet{Supponen:2017sw} for the maximum pressure for aspherical bubble collapses 
\begin{equation}
  p_\mathrm{max}\propto c_l \sqrt{\rho_l \Delta p} (R_0/d)^{1.25}= p^{\ast}(R_0/d)^{1.25}, 
  \label{eq:supponen_pmax} 
\end{equation}
where $d$ denotes the distance to the focus point. 

\section{Near-wall bubble collapses with varying stand-off distances $S/R_0$}
\label{sec:stand_off}

In this section, simulation results of collapsing bubbles for varying stand-off distances $S/R_0$ and a grid resolution of $240\,c/R_0$ are presented. \Cref{ss:SR0,ss:SRneg,ss:SRpos,ss:SRdet} analyze the collapse behavior of the different configuration (see also \cref{fig:setup}) and \cref{ss:comp} compares and discusses the induced pressure impacts and collapse behaviors. 

\subsection{Wall-attached bubble with zero stand-off distance $(S/R_0=0)$}
\label{ss:SR0}

\begin{figure}
  \centering
  \subfigure{\includegraphics[width=\linewidth]{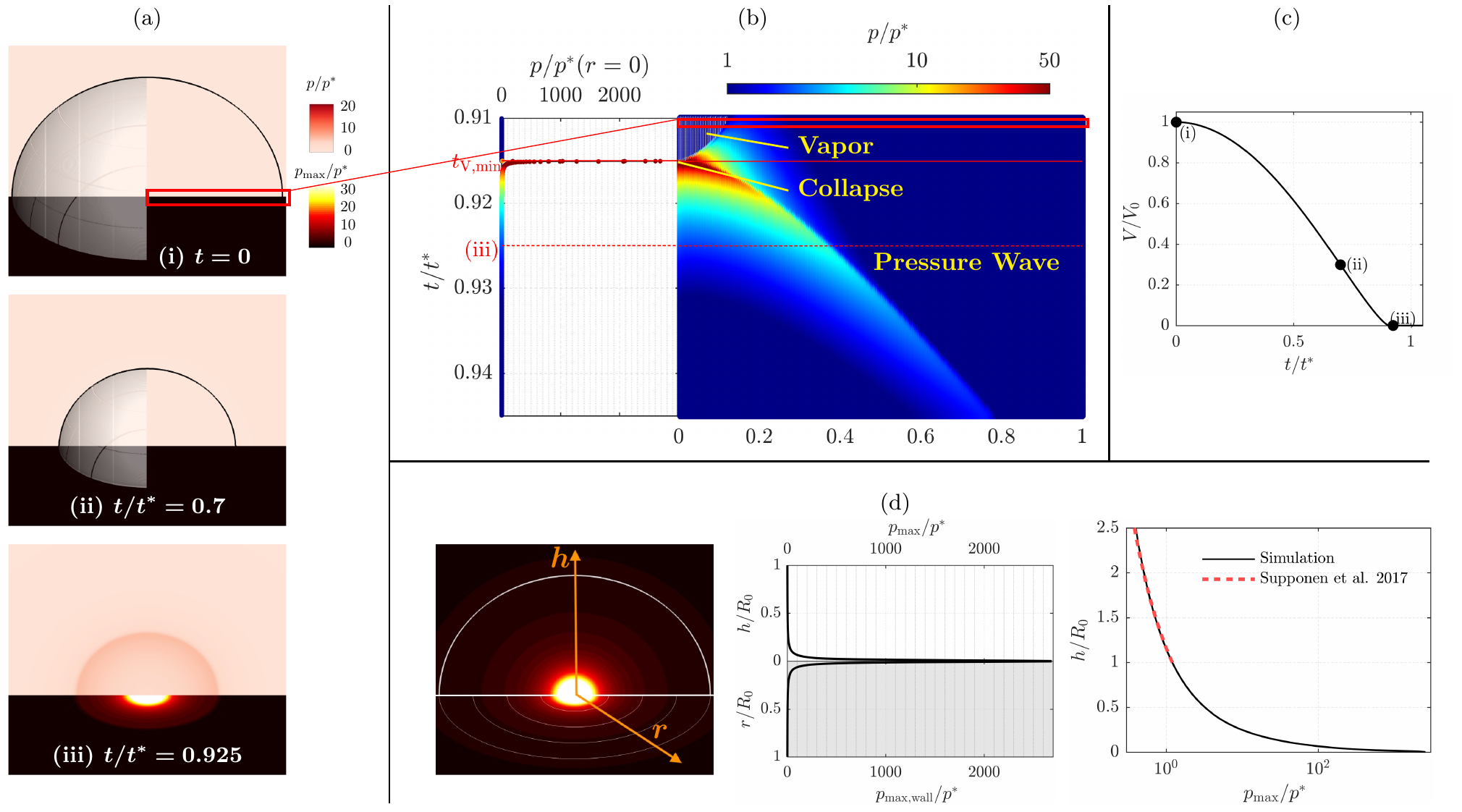}}
  \caption{Collapsing wall-attached bubble with $S/R_0=0$. 
  (a) Time series ((i)-(iii)) with instantaneous pressure ($p/p^{\ast}$) on the midplane with isoline/isosurface 10\% vapor and the maximum pressure at the wall ($p_\mathrm{max}/p^{\ast}$) (time increases from top to bottom). 
  (b) Wall-centered pressure signal ($r=0$) and a wall-pressure-diagram showing the instantaneous pressure in the first domain cell over time (time increases from top to bottom, $log(p/p^{\ast})$), time of minimum bubble volume $t_\mathrm{V,min}$ is indicated by a red solid line and depicted time steps by red dashed lines. 
  (c) Temporal evolution of the bubble volume, time steps shown in (a) are indicated with a dot.
  (d) Maximum pressure distribution with $p_\mathrm{max}/p^{\ast}$ in the domain (midplane with the initial bubble interface as a thick white line) and at the wall (white orientation rings at $r/R_0=\{0.25,\,0.5,\,0.75,\,1\}$) (right); extracted $p_\mathrm{max}/p^{\ast}$ in wall-normal $h$ and wall-parallel direction $r$ (center); extracted $p_\mathrm{max}/p^{\ast}$ in wall-normal direction $h$ compared with the empirical correlation of \citet{Supponen:2017sw} (see \cref{eq:supponen_pmax}) (right).}
  \label{fig:ts_s_r_0}
\end{figure}

\Cref{fig:ts_s_r_0}~(a) presents a time series for a collapsing wall-attached bubble with $S/R_0=0$. 
The pressure impact on the wall is shown as a wall-pressure-diagram in \cref{fig:ts_s_r_0}~(b), the temporal evolution of the bubble volume in \cref{fig:ts_s_r_0}~(c), and the maximum pressure distribution in \cref{fig:ts_s_r_0}~(d). 

The collapse of a wall-attached bubble with $S/R_0=0$ resembles that of a spherical bubble, see \cref{fig:ts_s_r_0}~(a,i--iii). Note that for a slip wall boundary condition, the collapse would be a spherical one. Although we apply no slip boundary conditions, inertia forces are predominant due to the high ambient pressure so that viscous effects become only relevant at the final stage of the collapse. The time series shows that the bubble interface accelerates uniformly and the retarding effect of the wall is annihilated by this dominance. The collapse occurs in the center of the bubble directly at the wall and results in the emission of a spherical pressure wave, decaying in intensity with the distance to the focus point (\cref{fig:ts_s_r_0}~(a,iii)). At the wall (\cref{fig:ts_s_r_0}~(a,i-iii)), we visualize the maximum recorded wall pressure, which can be interpreted as continuous monitoring of the wall deformation. To study the relevant mechanisms for potential wall damage in detail, we present a wall-pressure-diagram in \cref{fig:ts_s_r_0}~(b) showing the time-resolved pressure impact on the wall (first domain cell). At the beginning of the considered time interval, the remaining vapor bubble is visible. The vapor is further compressed by the surrounding liquid and the subsequent collapse ($t_\mathrm{V,min}$) causes the maximum wall pressure. After the collapse, the wall-pressure-diagram shows a radially outwards propagating pressure wave and its decay.

\Cref{fig:ts_s_r_0}~(d) visualizes the maximum pressure distribution. The extracted $p_\mathrm{max}$ in wall-normal direction (\cref{fig:ts_s_r_0}~(d), left) shows a very good agreement with the empirical correlation proposed by \citet{Supponen:2017sw} (see \cref{eq:supponen_pmax}). 

\subsection{Wall-attached bubble with a negative stand-off distance $(S/R_0<0)$}
\label{ss:SRneg}

\begin{figure}
  \centering
  \subfigure{\includegraphics[width=\linewidth]{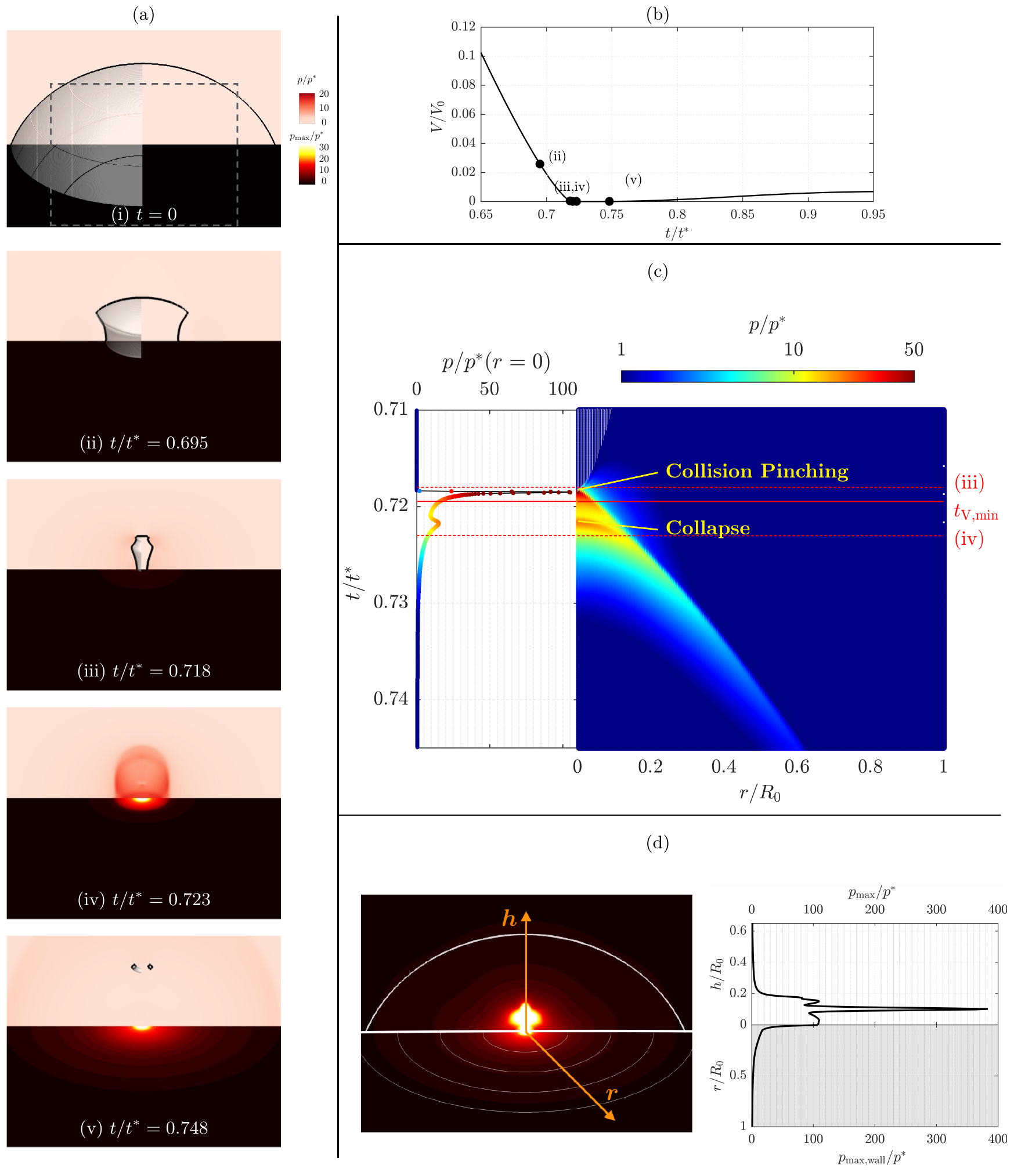}}
  \caption{Collapsing wall-attached bubble with $S/R_0=\text{--}0.35$. 
  (a) Time series (i)--(v) (in (ii)--(v) the indicated region in (i) is magnified); 
  (b) Temporal evolution of the bubble volume for selected time; 
  (c) Wall-pressure-diagram (see caption \cref{fig:ts_s_r_0}); 
  (d) Maximum pressure distribution (see caption \cref{fig:ts_s_r_0}). }
  \label{fig:ts_s_r_neg}
\end{figure}

\Cref{fig:ts_s_r_neg} shows the data for a collapsing wall-attached bubble with $S/R_0=\text{--}0.35$ consisting of a time series in~(a), the temporal evolution of the bubble volume in~(b), a wall-pressure-diagram in~(c), and the maximum pressure distribution in~(d). A schematic representation of the collapse and rebound behavior is visualized in \cref{fig:sketch}~(b). 

The wall-attached bubble is pinched at its maximum expansion in the circumferential direction, resulting in a mushroom shape (\cref{fig:ts_s_r_neg}~(a,ii,iii)), as also reported by \citet{Shima:1977df,Lauer:2012jh}. The radially inwards directed flow reaches very high velocities, exceeding $20\, u^{\ast}$. The subsequent collision of these inward moving liquid fronts induces a high pressure peak, which is visible in the wall-pressure-diagram and the wall-centered pressure signal in \cref{fig:ts_s_r_neg}~(c). Shortly afterwards, the remaining upper part (the ’mushroom head’) collapses and emits a pressure wave. When this wave reaches the wall, it causes the smaller second pressure increase (see \cref{fig:ts_s_r_neg}~(c)). Due to the conservation of momentum, the preceding radial inward flow at the pinching now causes an upward flow with velocities greater than $10\,u^{\ast}$. In the shear layer of this upward flow, the liquid evaporates, surrounding the flow with a vapor torus (see \cref{fig:ts_s_r_neg}~(a,v)). 

\Cref{fig:ts_s_r_neg}~(d) shows the distribution of the maximum pressure. As can be seen in the extracted data, the highest pressure occurs at the focus point of the collapse. The pressure peaks along the symmetry line and in the center of the wall are due to the collision of the liquid fronts. 

\FloatBarrier
\subsection{Wall-attached bubbles with positive stand-off distances $(0<S/R_0<1)$}
\label{ss:SRpos}

\begin{figure}
  \centering
  \subfigure{\includegraphics[height=0.9\textheight]{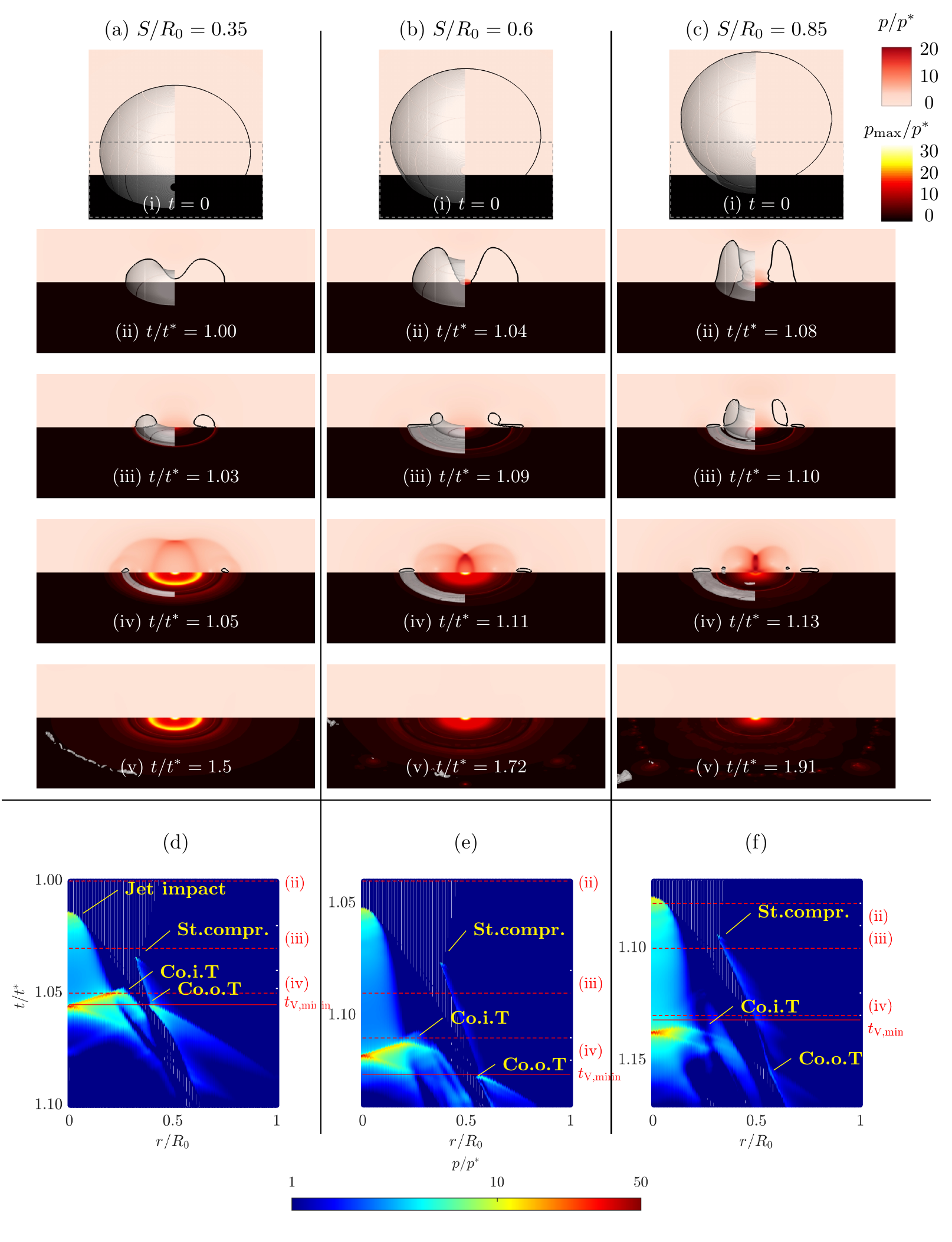}}
  \caption{Time series ((a)--(c)) and wall-pressure-diagrams ((d)--(f)) for wall-attached bubbles with $S/R_0 > 0$. In (a)--(c): (ii)--(v) are the magnified regions indicated in (i). Left column (a,d) for $S/R_0=0.35$, central column (b,e) for $S/R_0=0.6$, right column (c,f) for $S/R_0=0.85$. For more information see caption \cref{fig:ts_s_r_0}. 'St.compr.', 'Co.i.T', 'Co.o.T' indicate the pressure waves emitted at the stopped compression, at the collapse of the inner torus, and the collapse of the outer torus, respectively. }
  \label{fig:ts_s_r_lt_1}
\end{figure}

\begin{figure}
  \centering
  \subfigure{\includegraphics[width=\linewidth]{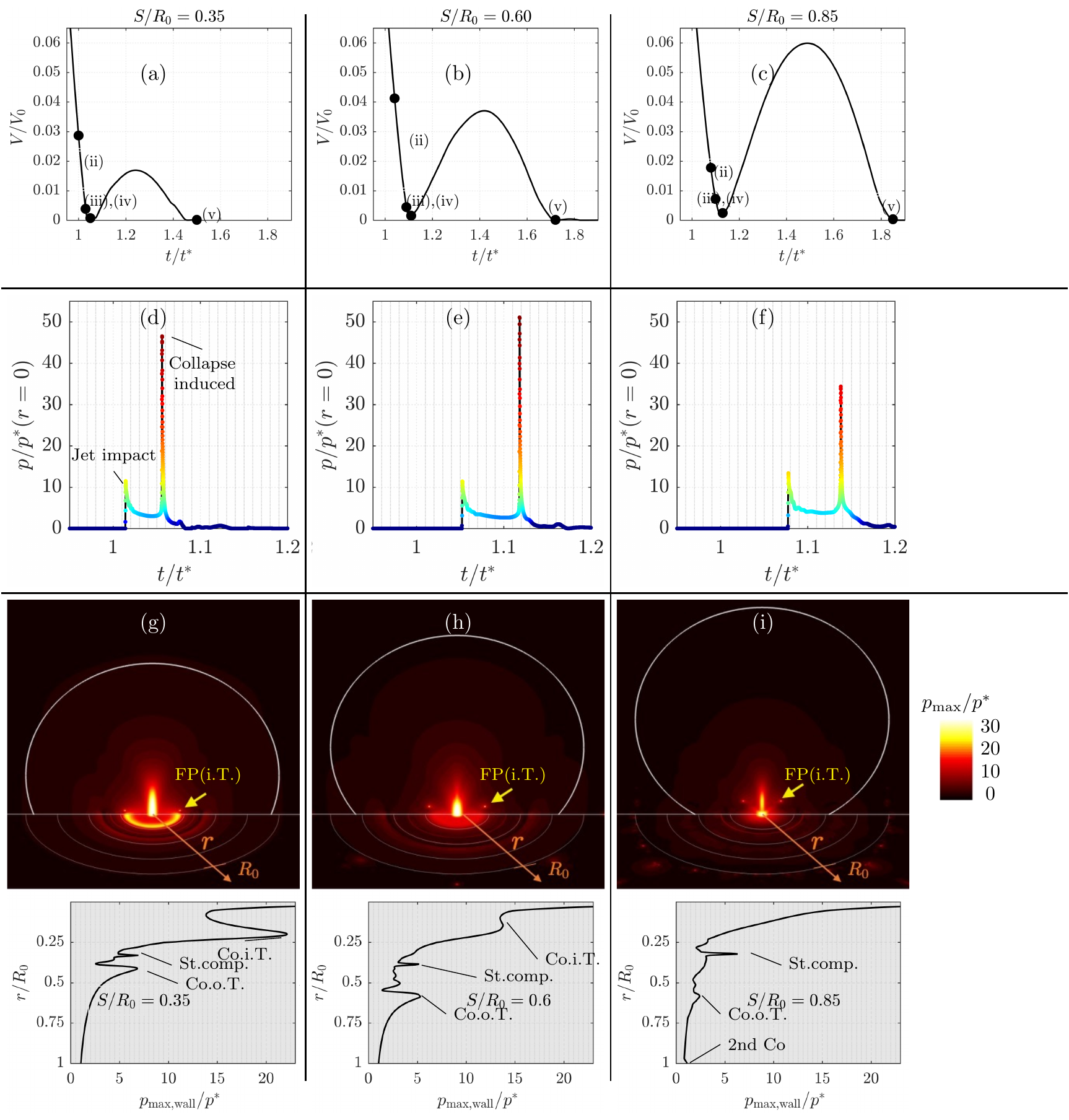}}
  \caption{Temporal evolution of the bubble volume ((a)--(c)) with marked time instants visualized in \cref{fig:ts_s_r_lt_1}~(a)--(c); wall-centered pressure signals ((d)--(f)); maximum pressure in the domain and at the wall and extracted maximum wall pressure in radial direction $r$ ((g)--(i)). In (g)--(i), diagrams are truncated for better visibility, maximum values see (d)--(e) and \cref{tab:co_data}. Left column (a,d,g) for $S/R_0=0.35$, central column (b,e,h) for $S/R_0=0.6$, right column (c,f,i) for $S/R_0=0.85$. 'St.compr.', 'Co.i.T', 'Co.o.T' indicate the pressure impact due to the stopped compression, the collapse of the inner torus, and the collapse of the outer torus, respectively. 'FP(i.T.)' marks the focus point of the collapsing inner torus. }
  \label{fig:ts_s_r_lt_1_void_wall}
\end{figure} 

\begin{figure}
  \centering
  \subfigure[]{\includegraphics[width=0.6\linewidth]{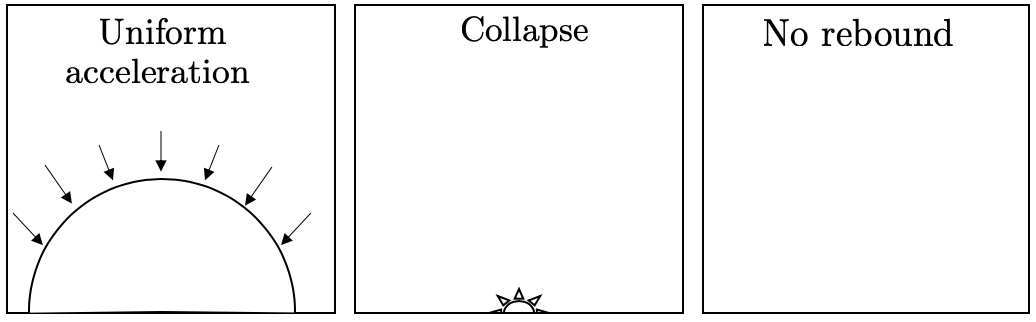}}
  \subfigure[]{\includegraphics[width=0.6\linewidth]{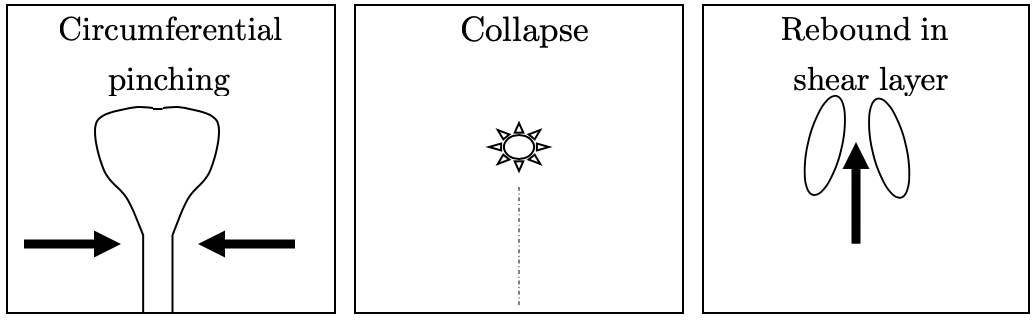}}
  \subfigure[]{\includegraphics[width=0.6\linewidth]{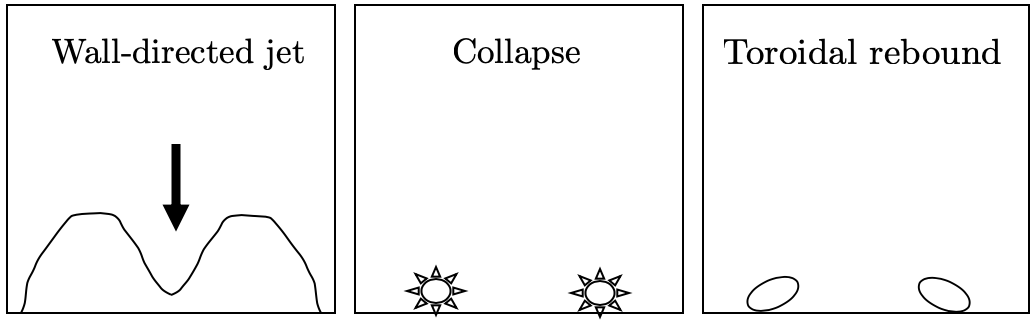}}
  \subfigure[]{\includegraphics[width=0.6\linewidth]{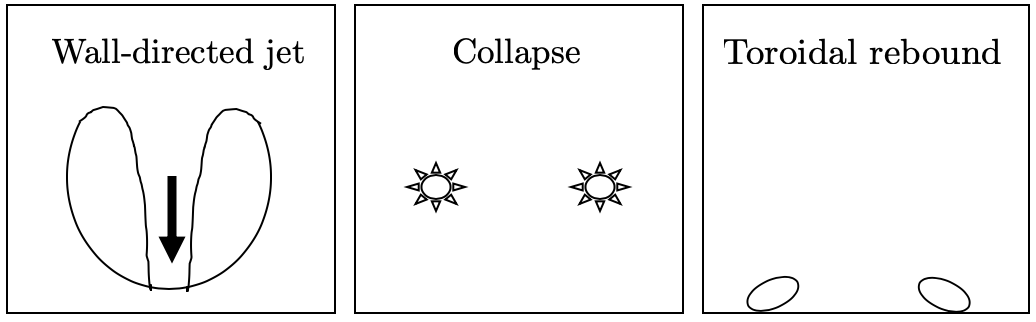}}
  \caption{Sketch of the collapse behavior of wall-attached bubbles with zero stand-off distance (a), a negative one (b), a positive one (c), and a wall-detached bubble (d). Adapted from \citep{Trummler:2020diss}.}
  \label{fig:sketch}
\end{figure}

In the following, we analyze the collapse behavior of wall-attached bubbles with a positive stand-off distance. The general behavior is well known and depicted in \cref{fig:sketch}~(b). \Cref{fig:ts_s_r_lt_1} shows time series for collapsing wall-attached bubbles in (a)--(c) and the corresponding wall-pressure-diagrams in (d)--(f). \Cref{fig:ts_s_r_lt_1_void_wall} contains the temporal evolution of the bubble volume in (a)--(c), the wall-centered pressure signals in (d)--(f) and the distribution of the maximum pressure in (g)--(i). 

For all cases, a wall-directed jet forms during the initial collapse phase resulting in a toroidal bubble. With increasing stand-off distance, the bubble is more elongated in wall-normal direction, see \cref{fig:ts_s_r_lt_1}~(a--c,ii). This observation was also reported by \citet{Philipp:1998eg}. When the jet hits the wall, it induces a pressure peak of about $10\,p/p^{\ast}$ and afterwards moves radially outwards along the wall pushing the vapor away (see \cref{fig:ts_s_r_lt_1}~(d)--(f) and \cref{fig:ts_s_r_lt_1_void_wall}~(d)--(f)). Consequently, the remaining torus is compressed from in- and outside. Then, the compression from outside is stopped, causing an outward propagating pressure wave and a first ring-shaped pressure impact on the wall (see \cref{fig:ts_s_r_lt_1}~(a--c,iii), (d)--(f) labeled with 'St.compr.'). Afterwards, the collapse of the torus becomes increasingly non-uniform with a portion near the wall being pinched away from the main torus, as was also observed by \citet{Lindau:2003oct,Trummler:2019ar}. At subsequent time steps, the bubble fragments into an inner bulged part and an outer pinched away one, which can be seen best in \cref{fig:ts_s_r_lt_1}~(c, iii). The inner part collapses first, emitting a pulse that travels radially inward, collides at the center, and continues in opposite direction (\cref{fig:ts_s_r_lt_1}~(a--c,iv)). 
These processes are also obvious in the wall-pressure diagrams (see \cref{fig:ts_s_r_lt_1}~(d)--(f) labeled with 'Co.i.T.'), especially at $S/R_0=0.35$. The collapse of the inner torus induces the largest wall pressure (see also pressure signals \cref{fig:ts_s_r_lt_1_void_wall}~(d)--(f)), which matches the findings by~\citet{Philipp:1998eg,Lauer:2012jh,Trummler:2019ar}. Then, the outer torus collapses emitting a less intense pressure wave propagating radially outwards, see wall-pressure-diagrams \cref{fig:ts_s_r_lt_1}~(d)--(f) 'Co.o.T'. Note that at the smaller stand-off distances ($S/R_0 = 0.35,0.6$), the collapse of the outer torus results in the minimum bubble volume $t_\mathrm{V,min}$, while at $S/R_0 = 0.85$ $t_\mathrm{V,min}$ is correlated with the collapse of the inner torus and $V_\mathrm{min}$ is slightly larger (\cref{fig:ts_s_r_lt_1_void_wall}~(c)). The subsequent rebound is toroidal and more pronounced at higher stand-off distances, see \cref{fig:ts_s_r_lt_1_void_wall} (a)--(c). At the second collapse, the torus fragments more and therefore the induced pressure impacts are not circumferentially uniform, which can be seen in \cref{fig:ts_s_r_lt_1}~(a--c,v). 

\Cref{fig:ts_s_r_lt_1_void_wall}~(g)--(i) visualizes the distribution of $p_\mathrm{max}/p^{\ast}$. Along the centerline in the domain and at the wall center, the superposition of pressure waves leads to high $p_\mathrm{max}$ values. Significant pressure peaks are also induced by the collapse of the inner torus (see 'FP(i.T.)'). At $S/R_0 = 0.35$, this peak is closest to the wall and thus the wall pressure increase underneath it, located at $0.2\,R_0$, is most significant. At $S/R_0 = 0.6$, that peak is attenuated and at $S/R_0 = 0.85$ no wall-pressure peak can be correlated to that collapse. The collapse of the outer torus causes a wall-pressure peak radially further outwards, which moves outwards and decreases with increasing $S/R_0$. Further, for all cases, there is a spike at about $0.3\,R_0$ associated with the abrupt stopping of the compression at the beginning (see \cref{fig:ts_s_r_lt_1}~(a--c,iii)). The second collapse only causes a visible wall pressure increase at $S/R_0 = 0.85$ at $r=R_0$. 

\FloatBarrier

\subsection{Wall-detached bubbles $(S/R_0>1)$}
\label{ss:SRdet}

\begin{figure}
  {\includegraphics[width=\linewidth]{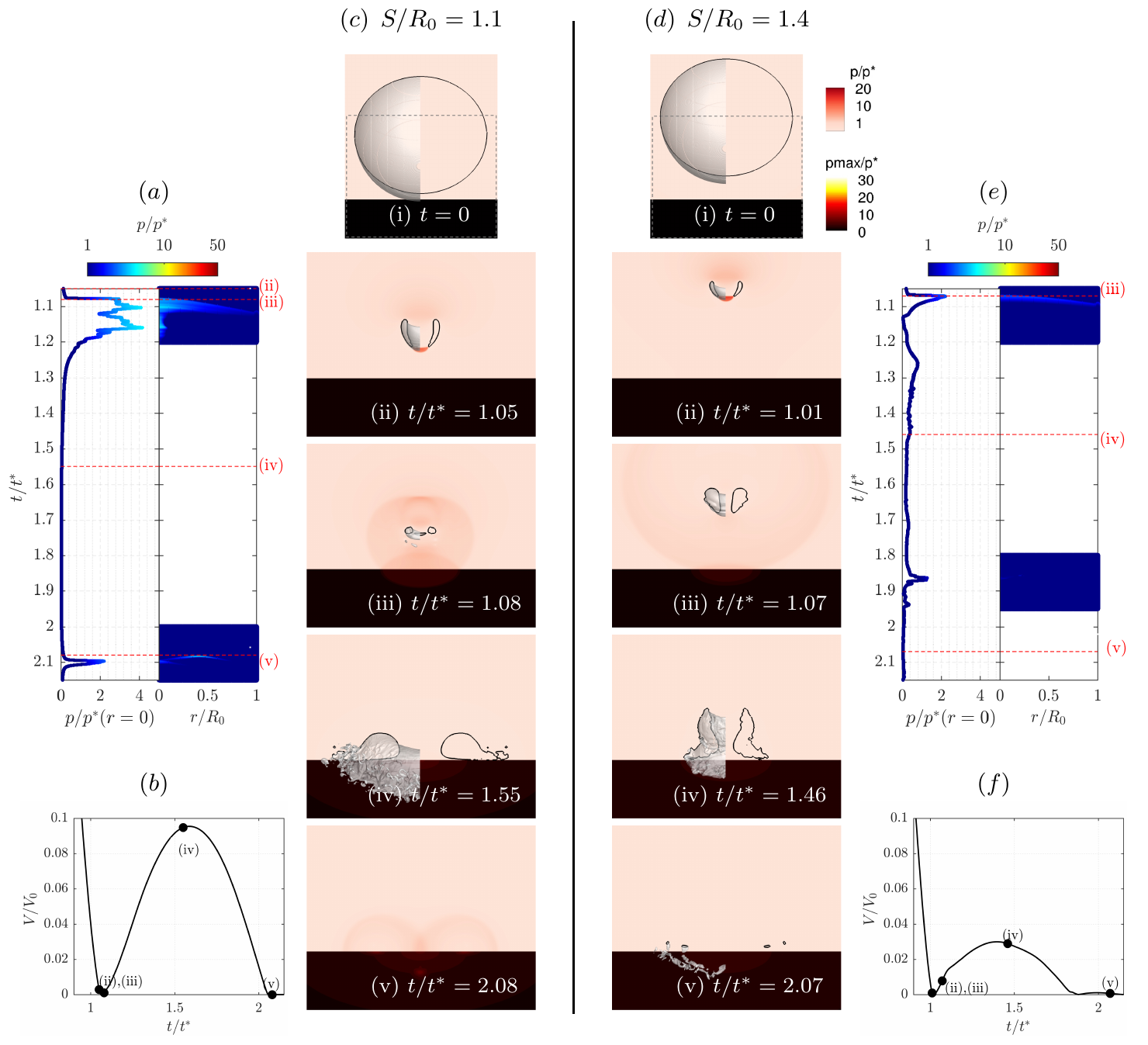}}
  \caption{Collapsing wall-detached bubbles with $S/R_0=1.1$ ((a)--(c)) and $S/R_0=1.4$ ((d)--(f)). Wall-pressure-diagrams (a,e), time series (c,d) ((ii)--(v) are the magnified regions indicated in (i)), and temporal evolution of the bubble volume (b,f). See also caption \cref{fig:ts_s_r_0}. }
  \label{fig:ts_s_r_gt_1}
\end{figure}

\begin{figure}
\center
  \subfigure{\includegraphics[trim={0 0 0 0},clip,width=\linewidth]{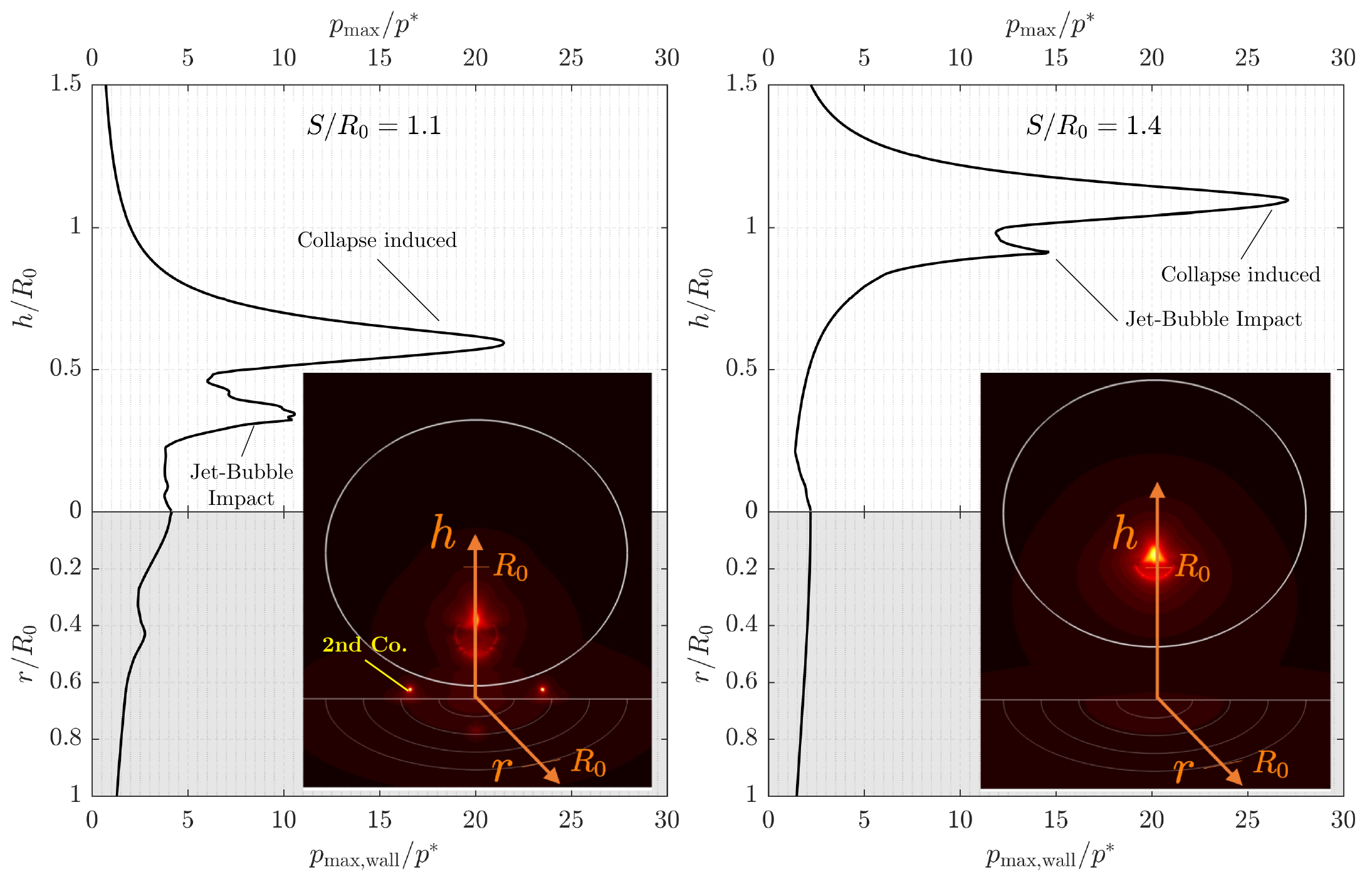}} 
  \caption{ Maximum pressure for $S/R_0=1.1$ (left) and $S/R_0=1.4$ (right). Extracted maximum pressure along the centerline in wall-normal direction $h$ and the wall in radial direction $r$, together with the visualization of the maximum pressure distribution. See also caption \cref{fig:ts_s_r_0}. }
  \label{fig:pmax_s_r_gt_1}
\end{figure}

\Cref{fig:ts_s_r_gt_1} shows time series~(c,d), wall-pressure-diagrams~(a,e) and temporal bubble-volume evolutions~(b,f) for collapsing wall-detached bubbles with $S/R_0=1.1$ and $S/R_0=1.4$. The maximum pressure distributions are shown in \cref{fig:pmax_s_r_gt_1}.

The aspherical pressure distribution leads to the formation of a jet penetrating the bubble. The jet is stopped at the lower bubble side, emitting a pressure wave (\cref{fig:ts_s_r_gt_1}~(c,d,ii)). At $S/R_0 = 1.4$, the intensity of the pressure wave is stronger, and the torus is more flattened and the bubble volume is smaller at jet impact. All three observations can be explained by the reduced anisotropy at the increased stand-off distance~\citep{Supponen:2016jna}. The pressure wave at jet-bubble-impact propagates radially outwards, hits the wall and is reflected there (\cref{fig:ts_s_r_gt_1}~(c,d,iv)). The impact on the wall is obvious in the wall-pressure-diagrams (\cref{fig:ts_s_r_gt_1}~(a,e)) and responsible for the highest peak at the wall center. At $S/R_0 = 1.1$, this wall-pressure increase is stronger since the jet-bubble-interaction is closer to the wall. 
Then, the first collapse takes place, followed by a rebound close to the wall (\cref{fig:ts_s_r_gt_1}~(c,d,iv)). At $S/R_0 = 1.1$ the rebound volume is larger (see \cref{fig:ts_s_r_gt_1}~(b,f)) and the second collapse results in a stronger impact on the wall (\cref{fig:ts_s_r_gt_1}~(a), \cref{fig:pmax_s_r_gt_1}). 

\Cref{fig:pmax_s_r_gt_1} shows the distribution of the maximum pressure and the extracted data. The induced pressure peak by the jet-bubble-impact can be clearly seen. These values are in a range of $\numrange{10}{15}\,p/p^{\ast}$, which corresponds to a jet velocity at impact of about $\numrange{10}{15}\,u/u^{\ast}$ (see \cref{eq:pwh}) and agrees with the findings of e.g.~\citet{Philipp:1998eg, Supponen:2016jna,Lauer:2012jh}. The highest maximum pressure in the domain is recorded on the centerline and caused by superposition of the pressure waves emitted at first collapse. At the wall, the highest maximum pressure occurs in the center and is due to the jet-bubble-impact (see \cref{fig:ts_s_r_gt_1}~(a,e)). At $S/R_0 = 1.1$, there is another peak at about $0.5\,R_0$ caused by the second collapse. 

\FloatBarrier

\begin{table}
\caption{Detailed data to the collapse behaviors and pressure impacts of all investigated stand-off distances. {\small(cfp)} indicates circumferential pinching and [1] refers to \citet{Supponen:2016jna}. S stands for 'spherical', T for 'toroidal', J for 'jet', and C for 'collapse'. See text for more information. }
\centering\begin{tabular}{|l|r||r||r|r|r||r|r||}
\hline
  $\boldsymbol{S/R_0}$ & \textbf{\text{--}0.35} & \textbf{0} & \textbf{0.35} & \textbf{0.6}& \textbf{0.85} &\textbf{1.1} &\textbf{1.4} \\
  \hline
  $\zeta=0.195 (S/R_0)^{-2}\;^{[1]}$      &    &   &   & 0.54 & 0.27& 0.16 &0.10 \\ 
  {Shape}       & S~  & S~ & T~  & T~ & T~  &T~  &T~  \\
   ${p_\mathrm{jet}}$    & {\footnotesize (cfp)} 107.20  & & 11.45  & 11.39 & 13.39  &10.45 &14.64\\
   ${p_\mathrm{jet,wall}}$ &  {\footnotesize (cfp)} 107.20 & - & 11.45  & 11.39 & 13.39  &4.20 &2.30\\
  ${p_c/p^{\ast}}$     & 383.50  & 2687.00 & 30.00  & 33.00 & 22.00   &10.00  &20.00  \\ 
  ${h_c/R_0}$     & 0.10 & 0.00 & 0.04& 0.07 & 0.12 &0.56 &0.94 \\
  ${r_c/R_0}$     &    &  & 0.22 & 0.22 & 0.15 &0.12 &0.07 \\ 
  ${r_{jet}/R_0}=0.5 \zeta^{2/3}\;^{[1]}$ 
             &    & &  & 0.33 & 0.21  &0.15 &0.11  \\
  {$V_\mathrm{reb}/V_0$}       & 0.01& 0.00& 0.02 & 0.04 & 0.06 & 0.10 & 0.03 \\
  ${p_\mathrm{max, wall}/p^*}$ & 107.20 & 2687.00 & 46.50& 51.40 & 33.70 & 4.20 &2.30\\
   {Mechanism $p_\mathrm{max, wall}$} &  J~  & C~ & C~  & C~ & C~  &J~  &J~  \\
\hline
  \end{tabular}
 \label{tab:co_data}
\end{table}

\subsection{Discussion and comparison} 
\label{ss:comp}

In the previous subsections, we observed distinctly different collapse and rebound behaviors for varying stand-off distances as illustrated in \cref{fig:sketch}. \Cref{tab:co_data} compares the bubble shape during collapse (spherical ($S$) or toroidal ($T$)), the jet-induced pressure peaks $p_\mathrm{jet}$, the collapse-induced pressure peaks $p_c$ (first collapse), the radial position $r_c$ and wall-normal position $h_c$ of the first collapse, the intensity of the rebound $V_\mathrm{reb}/V_0$, the maximum wall pressure $p_\mathrm{max,wall}$, and the mechanism causing the maximum wall pressure (jet ($J$) or collapse ($C$)). 

$S/R_0 \leq 0$ leads to a spherical collapse and $S/R_0>0$ to a toroidal one, where the torus is created by a wall-normal jet. When this jet hits the wall or the distal bubble side, it is abruptly stopped and induces a pressure peak ($p_\mathrm{jet}$). The range of the estimated jet velocities $u_\mathrm{jet}/p^{\ast}\approx p_\mathrm{jet}/p^{\ast}$ (\cref{eq:pwh}) agrees with data from the literature~\citep{Supponen:2016jna, Philipp:1998eg,Plesset:1971hu, brujan2002final}. The increasing jet velocity with increasing stand-off for $S/R_0 > 1$ matches the findings by \citet{Supponen:2016jna, Philipp:1998eg}. With respect to the impact of the jet on the wall ($p_\mathrm{jet,wall}$) our simulations are in very good agreement with the data of \citet{Philipp:1998eg}. At $S/R_0<0$ a circumferential pinching takes place, which can also be interpreted as a lateral wall-parallel jet~\citep{Lauer:2012jh}. The collision of the fronts induces high peak values, here counted as $p_\mathrm{jet}$ and $p_\mathrm{jet,wall}$. 

The collapse-induced pressure peak for spherical collapses with $p_c/p^{\ast} = \numrange{400}{3000}$ is significantly higher than for toroidal ones with $p_c/p^{\ast} =\numrange{10}{30}$. $p_c$ decays with increasing $S/R_0$ for $0 \leq S/R_0 <1 $, and then increases again for $S/R_0 >1 $. This trend matches experimental findings for the intensity of the emitted pressure wave \citep{Vogel:1988er, Vogel:1989jfm, Supponen:2017sw}. 

In addition to the intensity, the position of the collapse is also an interesting parameter. The radius of the collapsing torus $r_c$ decreases with increasing $S/R_0$. \citet{Supponen:2016jna} found that the jet impact diameter $d_\mathrm{jet}$ correlates with the anisotropy parameter $\zeta$ as $d_\mathrm{jet} = R_0 \zeta ^{2/3}$, where $\zeta$ reads for walls $\zeta=0.195 (S/R_0)^{-2}$. Assuming $r_c \approx 0.5\,d_\mathrm{jet}$, we can compare her findings with our data and observe the same trend, only that our values are slightly smaller (see \cref{tab:co_data}). Moreover, our data for $r_c$ roughly match experimentally obtained damage radii of wall-attached bubbles of about $0.33\,R_0$~\citep{Philipp:1998eg}. The distance of the collapsing torus to the wall $h_c$ increases, as expected, with $S/R_0$. 

The maximum wall pressure $p_\mathrm{max, wall}$ depends on the intensity of the highest pressure and its relative position to the wall. At $S/R_0 = 0$, the collapse takes place directly at the wall with an extremely high collapse-induced pressure peak, resulting in the highest $p_\mathrm{max, wall}$ for all $S/R_0$. At $S/R_0 =\text{--}0.35$, $p_\mathrm{max, wall}$ is induced by the collision of the circumferential jet and one order of magnitude smaller than at $S/R_0=0$. For $S/R_0 > 0$, $p_\mathrm{max, wall}$ decreases with increasing $S/R_0$, is collapse-induced for wall-attached bubbles and jet-induced for wall-detached ones, and is two orders of magnitude smaller than at $S/R_0=0$. The observed decrease of the maximum wall pressure with increasing stand-off distance matches experimental observations for wall-attached bubbles at atmospheric conditions~\citep{Shima:1983we, Shima:1984vw, Tomita:1986gy} and is consistent with measured cavitation damage depths~\citep{Philipp:1998eg}. Further, our data is in an excellent agreement with numerically predicted maximum wall pressures for similar configurations~\citep{Lauer:2012jh,Trummler:2019ar}. $p_\mathrm{max, wall}$ as a function of $S/R_0$ is also further discussed at the end of the next section. 

Here we consider generic configurations of spherical bubbles that are cut or detached from the wall at different positions. In recent studies \citep{Lechner:2019fp, Lechner:2020jet}, the growth phase of the bubbles was included, resulting in elliptical bubble shapes at maximum expansion for small stand-off distances. Additionally, surface tension was also taken into account. These studies provided new insights into the subsequent deformation of the bubble during collapse, leading to the formation of high velocity jets that can potentially induce high pressure peaks on the wall.

\FloatBarrier
\section{Effect of the grid resolution} \label{sec:grid}

 \begin{figure}
  \centering
  \subfigure{\includegraphics[width=\linewidth]{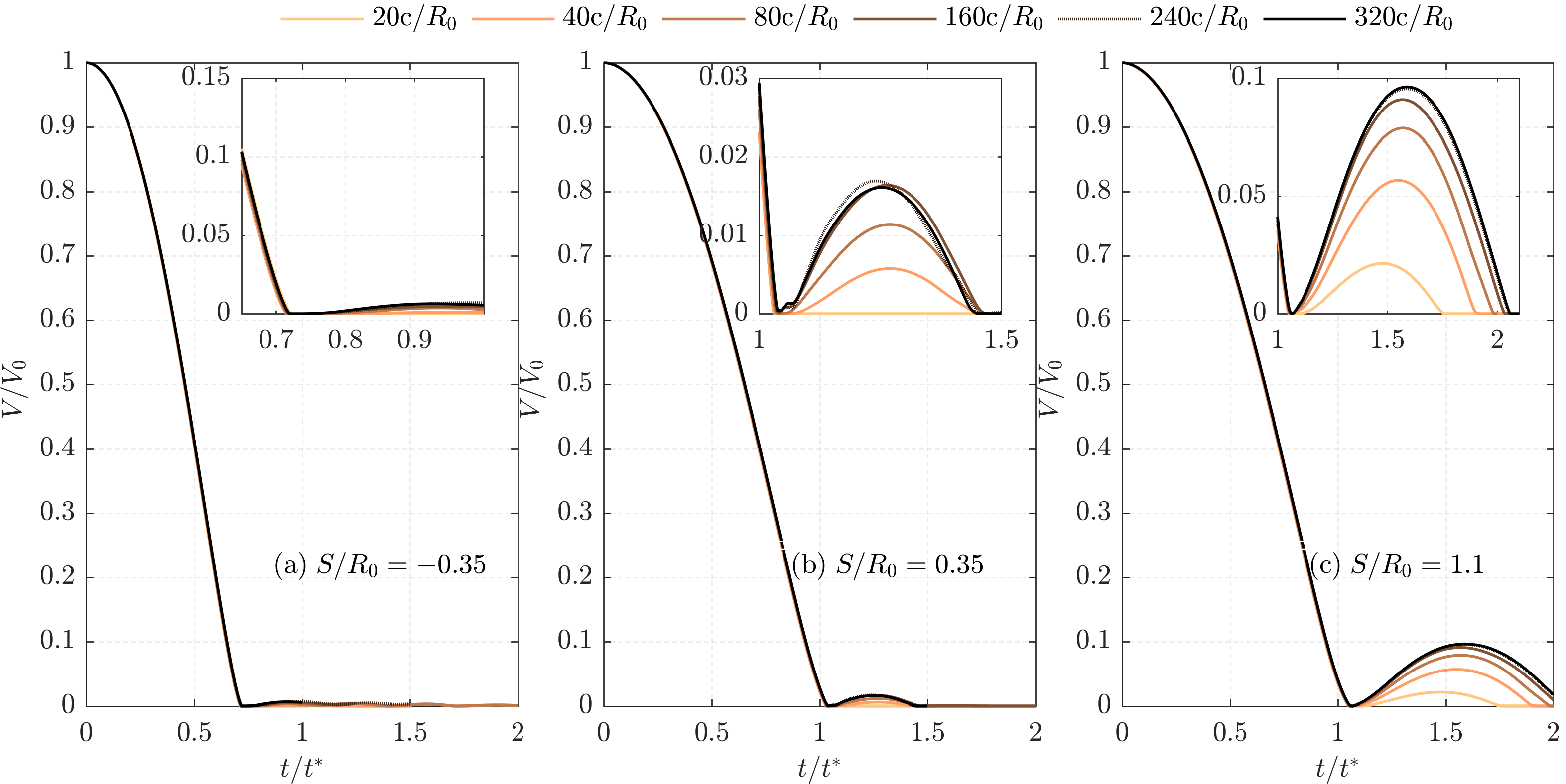}}
  \caption{Temporal evolution of the bubble volume for different grid resolutions of $20\,c/R_0$ to $320\,c/R_0$.\newline (a)$\;S/R_0=\text{--}0.35$, (b)$\;S/R_0=0.35$, (c)$\;S/R_0=1.1$. Since $240\,c/R_0$ does not fit into the sequence with 1:2 refinements, the data are shown with a dotted line.}
  \label{fig:grid_rebound}
\end{figure}

 \begin{figure}
 \centering
  \hspace{1cm}
  \subfigure{\includegraphics[width=0.8\linewidth]{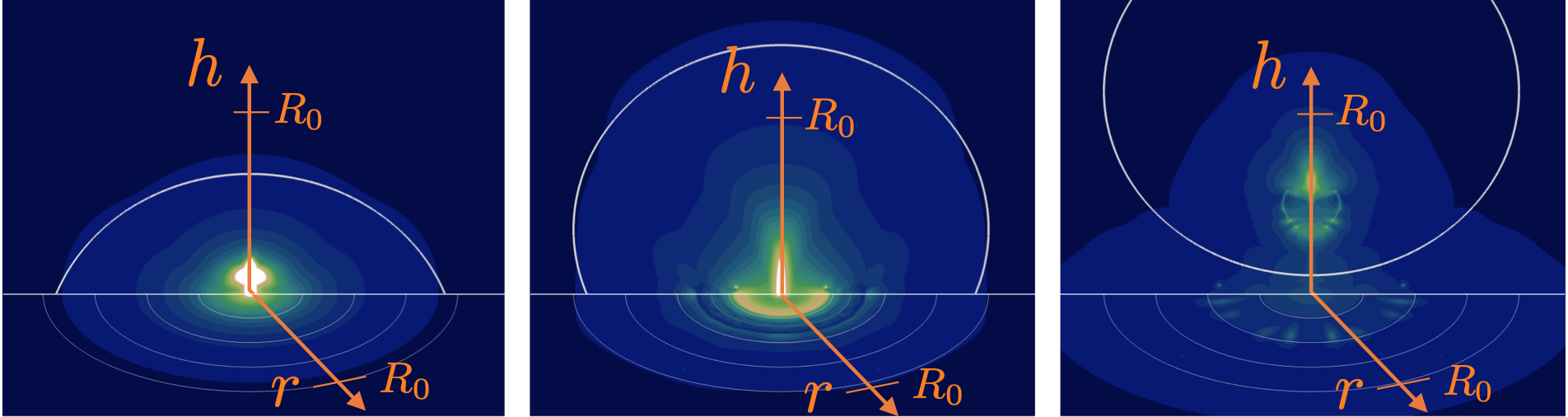}}
  \subfigure{\includegraphics[width=0.08\linewidth]{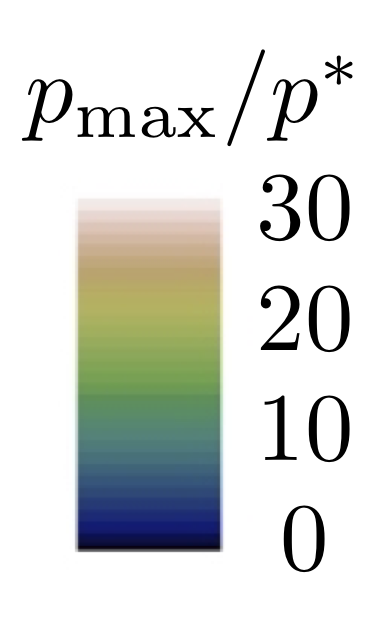}}
  \subfigure{\includegraphics[width=\linewidth]{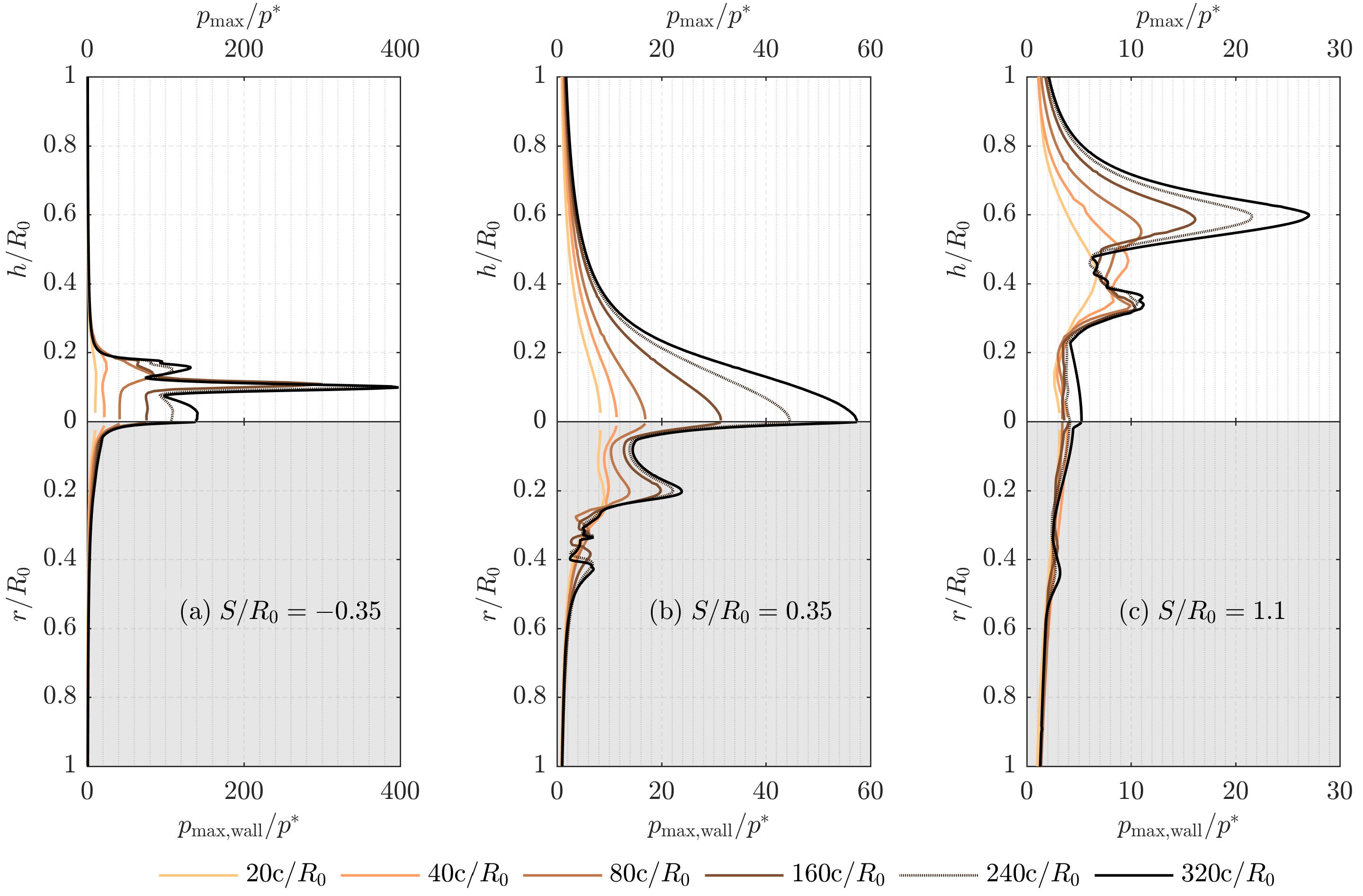}}
  \caption{Extracted data of maximum pressure in the domain and at the wall for different grid resolutions of $20\,c/R_0$ to $320\,c/R_0$. (a)$\;S/R_0=\text{--}0.35$, (b)$\;S/R_0=0.35$, (c)$\;S/R_0=1.1$. Since $240\,c/R_0$ does not fit into the sequence with 1:2 refinements, the data are shown with a dotted line.}
  \label{fig:grid_pmax_extracted}
 \end{figure}

 \begin{figure}
  \centering
  \subfigure{\includegraphics[height=19cm]{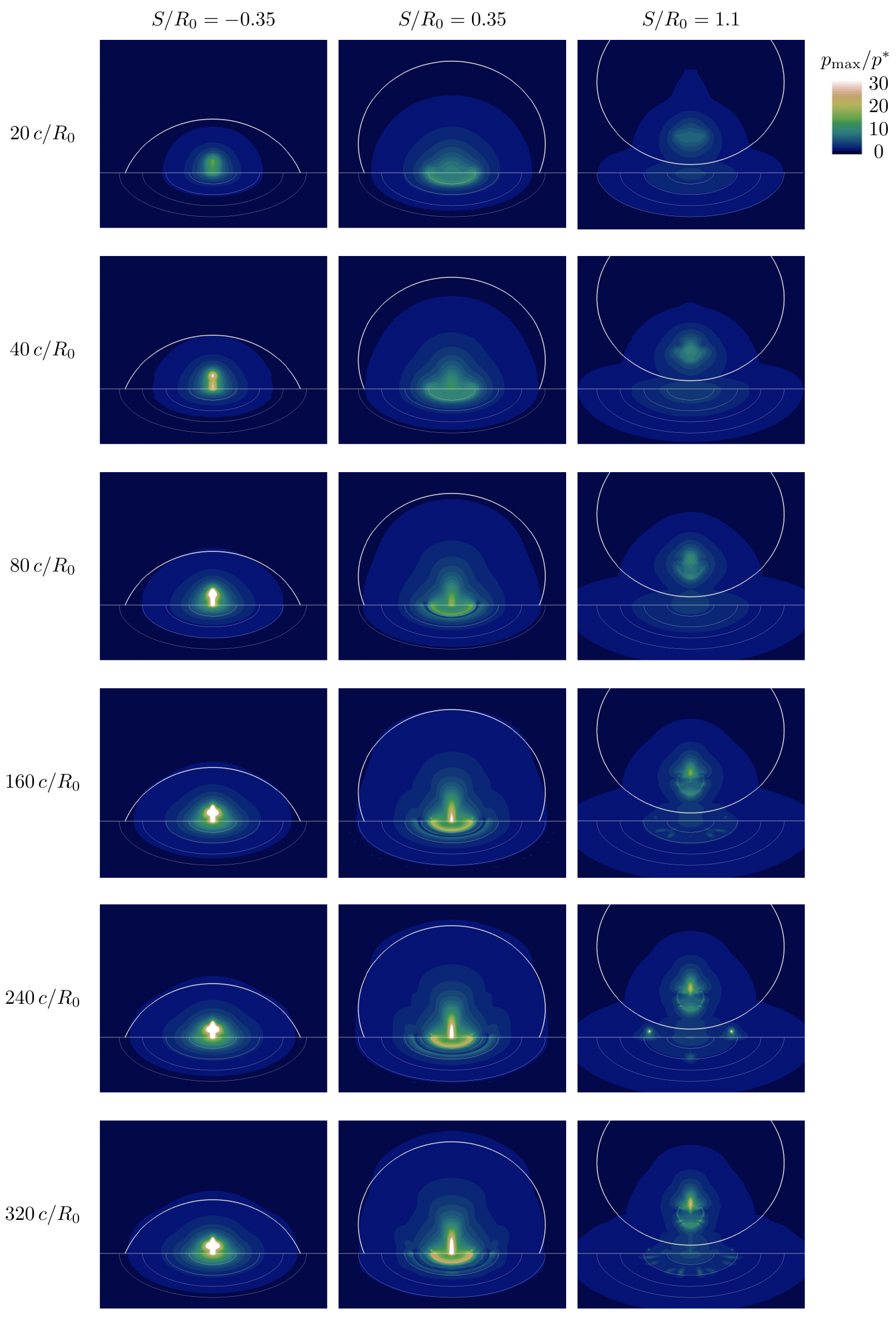}}
  \caption{Maximum pressure on the domain midplane and the wall for different grid resolutions $N_c/R_0$ (columns) and stand-off distances $S/R_0$ (rows). On the midplane the initial bubble interface is shown as a thick white line and at the wall are white rings at $r/R_0=\{0.25,\,0.5,\,0.75,\,1\}$.}
  \label{fig:grid_pmax_vis}
\end{figure}

 \begin{figure}
  \centering
  \subfigure{\includegraphics[width=0.8\linewidth]{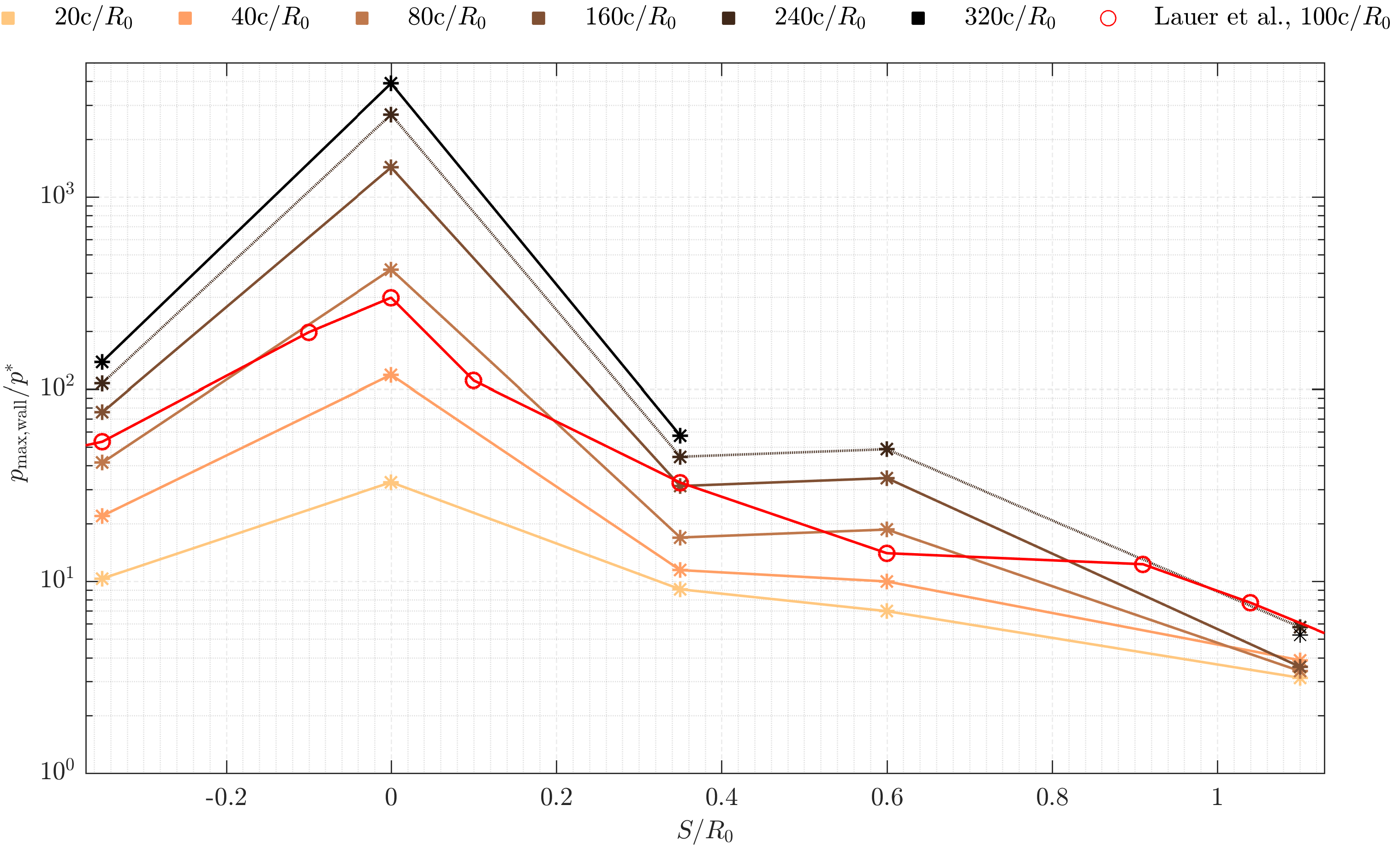}}
  \caption{Maximum pressure recorded at the wall $p_\mathrm{max,wall}$ for varying stand-off distances $S/R_0$ and different grid resolutions. Results obtained by \citet{Lauer:2012jh} are shown for comparison.}
  \label{fig:grid_pmax_wall}
\end{figure}

The previous section has shown that there are different collapse behaviors, rebound intensities, and maximum pressure distributions for varying stand-off distances. In the following, we assess the effect of the grid resolution on these features. For this purpose, we have performed simulations with grid resolutions of $20,\,40,\,80,\,160,\,240,$ and $320\,c/R_0$. 
Note that $240\,c/R_0$ does not follow the otherwise present 1:2 refinement. The simulations with $320\,c/R_0$ are that resource-intensive and produce such large amounts of data that the detailed analysis in the previous section would not have been feasible with that resolution, motivating the resolution of $240\,c/R_0$.

\Cref{fig:grid_rebound} presents the temporal evolution of the bubble volume for different grid resolutions and stand-off distances. In our simulations, the time of minimum bubble volume $t_\mathrm{V,min}$ is well captured for all grid resolutions, while the rebound is clearly grid dependent and converges for $N_c/R_0 \to 300$. 

The peak pressures in the domain and at the wall are also affected by the grid resolution. \Cref{fig:grid_pmax_extracted} shows the extracted maximum pressure on the centerline in the domain and the circumferentially averaged values at the wall and \Cref{fig:grid_pmax_vis} visualizes the maximum pressure distribution for the different grid resolutions. 

At $S/R_0=\text{--}0.35$, there is a circumferential pinching of the bubble and a spherical collapse of the upper part with an offset to the wall, see \cref{ss:SRneg}. With increasing grid resolution, the focus point of the collapse moves closer to the wall. For $N_c/R_0 \geq 160$, the wall-normal position of the focus point remains the same and only the peak value still increases. On the centerline beneath the focus point and directly at the wall, the maximum pressure is induced by the collision of the liquid fronts at circumferential pinching. Due to the acceleration of the liquid radially inwards, the smallest distance to the centerline, and thus the cell length, determines the maximum velocity and the induced peak pressure. For $N_c/R_0 \leq 160$, the peak value at the wall center increases inversely proportional to the cell length. At higher resolutions this proportionality is no longer given, possibly due to viscous effects. 

At $S/R_0=0.35$, the first toroidal collapse induces the maximum pressure, see \cref{ss:SRpos}. The ring-shaped pressure impact on the wall caused by this collapse is already visible at $20\,c/R_0$ and more pronounced at higher grid resolutions without changing its radial position ($\approx 0.2\,R_0$). At this stand-off distance, three radial pressure peaks occur in the maximum pressure distribution at the wall. As discussed in \cref{ss:SRpos}, these are induced by the stopped compression from outside (spike at $\approx 0.3\,R_0$), the collapse of the inner torus ($\approx 0.2\,R_0$) (the torus fragments into an inner and an outer one) and the collapse of the outer torus ($\approx 0.4\,R_0$). For $N_c/R_0 \geq 160$, all three peaks are recorded and the wall pressure distribution converges, except for the peak value in the wall-center. 

At the detached configuration ($S/R_0=1.1$), a wall-directed jet pierces the bubble and a pressure peak is induced when the jet hits the distal bubble side. The subsequent collapse of the torus induces peak values on the centerline at $h\approx0.6\,R_0$. After the rebound, the second collapse takes place near the wall and induces a pressure impact on the wall at $r\approx 0.5\,R_0$, see \cref{ss:SRdet}. The pressure peak caused by the jet-bubble-impact is visible for all grid resolutions except for the coarsest grid. For $N_c/R_0\geq80$, this peak converges to $10\, p^{\ast}$, which coincides with the expected value (see \cref{ss:SRdet}). The collapse-induced peak on the centerline is again strongly grid-dependent. At the wall, the jet-bubble-impact induces a pressure peak in the center with little grid dependence. The ring-shaped pressure impact of the second collapse ($\approx 0.5\,R_0$) is captured for $N_c/R_0 \geq 80$. Its intensity and distribution depends on the grid resolution and grid orientation and is therefore not radially symmetric.

\Cref{fig:grid_pmax_wall} shows the maximum wall pressure ${p}_\mathrm{max, wall}$ for different grid resolutions and stand-off distances $S/R_0$. 
At all configurations, the maximum pressure was recorded in the center.
With an increasing absolute stand-off distance $|S/R_0|$, ${p}_\mathrm{max,wall}$ decreases. This matches experimental and numerical investigations from the literature, see \cref{ss:comp}. Our data for $80\,c/R_0$ are in good agreement with the reference data of \citet{Lauer:2012jh} with $100\,c/R_0$. Regarding grid dependence, we observe the following. For wall-attached bubbles, the increase in ${p}_\mathrm{max,wall}$ is approximately inversely proportional to the cell length, whereas for wall-detached bubbles ${p}_\mathrm{max,wall}$ is less grid dependent. Overall, the relative ratio of pressure peaks is comparable at identical grid resolutions and, consequently, the data obtained at a certain grid resolution allow for a comparative and qualitative assessment of the cavitation erosion potential.

In summary, the most important collapse features are already captured at the coarsest grid resolution of $20\,c/R_0$. The main features of the maximum pressure distribution, such as pronounced pressure peaks at toroidal collapses, are mostly resolved for $N_c/R_0\geq 80$ and at higher resolutions mainly the intensity of the peaks changes. For the peak values, we observe that the values induced by a collision of accelerated liquid fronts or a collapse are approximately inversely proportional to the cell length, as also found by \citet{Mihatsch:2015db}. Away from the focus point, the grid dependence of the maximum pressure vanishes, which is consistent with the observations of \citet{schmidt2014assessment,Mihatsch:2017diss}. 
For the rebound and the second collapse we find a strong grid dependence, but a convergence of the rebound volume for high grid resolutions. 

A doubling of the grid resolution (1:2 refinement) leads for 3-D calculations to 8-times as many cells. Additionally, the average time step is inversely proportional to the cell length, thus doubling the number of iterations. Hence, a 1:2 grid refinement requires 16-times as many resources for an assumed ideal scaling. In fact, a finer grid requires even more resources, because the computing domain is typically distributed over more compute cores, resulting in additional communication overhead between the cores. This is the reason why quantitative estimates for resolution effects are pivotal for engineering predictions that typically are marginally resolved.

\section{Conclusion}\label{sec:Conclusion}

We have numerically investigated the collapse and rebound behavior and the erosion potential of vapor bubbles collapsing close to a wall. Contrary to most previous studies, we consider phase transition and thus also capture rebound processes. We have evaluated the maximum pressure in the domain and at the wall and additionally presented time-resolved wall pressure signals, which allow for the identification of the mechanism inducing the highest wall pressure peaks. For bubbles with a negative stand-off distance, the collision of the circumferential pinching induces the highest peak pressure at the wall. For wall-attached bubbles with a positive stand-off distance, the first collapse leads to the highest wall pressures, which are associated with the superposition of the emitted pressure waves in the center. For wall-detached bubbles, the highest pressure at the wall is induced by the pressure wave emitted at jet-bubble-impact and the second collapse can also cause a relevant increase in wall pressure. 

Furthermore, the influence of the grid resolution on the collapse dynamics and the maximum pressures distribution was investigated. For this purpose, we have performed simulations of collapsing bubbles at different stand-off distances with grid resolutions from $20$ to $320\,c/R_0$. The grid study shows that the collapse behavior and the maximum pressure distribution are already captured with the coarsest resolution. With increasing resolution, the rebound and characteristic features of the maximum pressure distribution are better resolved and converge at higher resolutions ($N_c/R_0\geq160$), except for the intensity of the peak values. In agreement with the literature, we found that peak values induced by a collision of accelerated liquid fronts scale approximately inversely proportional to the cell length. The presented comprehensive study provides a reference for choosing the required grid resolution at future simulations. 

We were able to identify the decisive processes for wall pressure peaks induced by near-wall bubble collapses under high ambient pressure. Already at the considered ambient pressure of 100 bar, the rebound volume was up to 10\% of the initial volume and significant wall pressure peaks were caused by the second collapse, which is in accordance with experimental observations that the second collapse can be decisive for cavitation erosion. In future studies, we plan to investigate the influence of the driving pressure difference on the collapse dynamics and erosion potential and to perform simulations under atmospheric conditions. This is motivated by the fact that at lower ambient pressures a stronger rebound~\citep{Tinguely:2012wo} and thus higher pressure peaks at the second collapse are expected. Additionally, a better comparability with experimental studies would be given, since these are mainly conducted under atmospheric conditions~\citep{Philipp:1998eg,dular2019high}. First numerical studies under atmospheric conditions were already presented in \citet{Trummler:2020diss}. 

\section*{Declaration of Competing Interest}

The authors declare that they have no known competing financial interests or personal relationships that could have appeared to influence the work reported in this paper.

\section*{Acknowledgment}

The authors gratefully acknowledge the Gauss Centre for Supercomputing e.V. (www.gauss-centre.eu) for funding this project by providing computing time on the GCS Supercomputers SuperMUC and SuperMUC-NG at Leibniz Supercomputing Centre (www.lrz.de). N.A.A. acknowledges support through the ERC Advanced Grant NANOSHOCK (grant agreement No. 667483). 

\section*{References}
\bibliography{main}
\bibliographystyle{elsarticle-harv}
\biboptions{authoryear}

\end{document}